\documentclass[amssymb,prd,aps,showpacs,nofootinbib,preprint]{revtex4-1}
\usepackage{latexsym}
\usepackage{graphicx,color}
\usepackage{graphicx}
\usepackage{amsmath,amssymb}
\usepackage{amsmath}
\usepackage[T1]{fontenc}
\usepackage[hang,small,up]{caption}
\newcommand{\sgn}{\mathop{\mathrm{sgn}}}
\makeindex
\begin{document}
\title{Thermodynamics aspects of noise-induced phase synchronization}
\author{Pedro D. Pinto$^{1}$}\author{Fernando A.
Oliveira$^{2,3}$}\email{fao@fis.unb.br}\author{Andr\'e L.A. Penna$^{2,3}$}
\affiliation{Universidade Federal do Oeste da Bahia, CP 47850-000, BA, Brazil,$^{1}$}
\affiliation{Instituto de F\'{i}sica, Universidade de Bras\'{i}lia, Brazil$^{2}$\\
International Center for Condensed Matter Physics\\CP 04455,
70919-970 Bras\'{i}lia DF, Brazil,$^{3}$}


\begin{abstract}

In this article, we present an approach for the thermodynamics of phase oscillators induced by an internal multiplicative noise. We analytically derive the free energy, entropy, internal energy, and specific heat. In this framework, the formulation of the first law of thermodynamics requires the definition of a synchronization field acting on the phase oscillators. By introducing the synchronization field, we have consistently obtained the susceptibility and analyzed its behavior. This allows us to characterize distinct phases in the system, which we have denoted as synchronized and parasynchronized phases, in analogy with magnetism. The system also shows a rich complex behavior, exhibiting ideal gas characteristics for low temperatures and susceptibility anomalies that are similar to those present in complex fluids such as water.

\end{abstract}
\date{\today }
\maketitle

\section{Introduction}

The synchronization of nonlinear oscillators is an important cooperative phenomenon widely applied in different disciplines, ranging from physics to social sciences \cite{Strogatz,Longa,Ciesta,Morgado,Acebron,Hong,Bonilla,Sonnenschein,Park,Reimann,Yu}. The advent of the phase reduction method has allowed an advantage in getting simpler equations for the study of limit cycle oscillators \cite{Kuramoto1,Nakao}, enabling a breakthrough in understanding the application of the synchronization phenomenon. It is recognized nowadays that the principal cause of synchronization in physical systems is due to the nonlinear coupling between the oscillators. Nevertheless, synchronization can also occur in decoupled oscillators through a phenomenon known as noise-induced synchronization. Indeed this phenomenon is found for a general class of limit cycle decoupled oscillators, where the occurrence of negative Lyapunov exponents is observed for sufficiently weak additive noise \cite{Zhou,Teramae,Pikovsky2,Pikovsky}. In coupled oscillators, it was observed that the common additive noise allows a reduction of the critical coupling, which leads to synchronization \cite{Nagai,Lai}.

In this scenario, a phenomenon rarely addressed in phase oscillators is the internal noise-induced synchronization, i.e., when synchronization is modulated by the state of the oscillators due to the multiplicative noise. This is a phenomenon that occurs in neural systems and is known as intrinsic coherent resonance \cite{Hanggi}. Additionally, another field that is still open is the thermodynamics to the noise-induced phase synchronization. Since these oscillators are inherently non-Hamiltonian systems, their approach occurs predominantly in the geometric point of view, i.e., through bifurcation analysis, applying the center manifold theory \cite{Crawford}. Indeed, only recently have emerged approaches that effectively consider aspects of the statistical thermodynamics of equilibrium and non-equilibrium \cite{Gupta} as well as those within the stochastic thermodynamics \cite{Sasa}.

In this article, we propose a way to construct the equilibrium thermodynamics of phase synchronization for oscillators governed by an internal multiplicative noise. This consists in extending the conventional Kuramoto--Sakaguchi model \cite{Sakaguchi} by including a phase-dependent multiplicative noise. From this, we derive the Fokker--Planck dynamics, where we show that the system sufficiently relaxes for thermodynamic equilibrium. This allows us to exactly determine the stationary phase density, order parameter, and temperature. From these quantities, we formulate the first law of thermodynamics that connects the internal energy and entropy with the concept of the synchronization field, which drives the synchronization of the system. Using the Legendre transform, we express the first law in terms of free energy. Thermodynamics is then constructed from the free energy, where expressions of entropy, internal energy, specific heat, and synchronization field are analytically obtained.

In fact, one of the reasons for the difficulty in establishing the full thermodynamics of phase oscillators is the absence in the literature of a synchronization field formulation. This is crucial to know the response of the system under the action of the internal multiplicative noise. From the synchronization field, we define susceptibility and analyze its behavior on the system. For a non-null order parameter, the existence of two phases is identified, which we call synchronized and parasynchronized phases, in analogy to magnetism. Susceptibility also shows us that the synchronized phase exhibits an anomalous region very similar to the region of anomalous behavior in water.

\section{The model}

We begin by introducing the Ito stochastic differential equation \cite{Gardiner} for phase oscillators whose dynamics governed by
\begin{equation}
\label{eq:kura_model}
\dot{\theta_i}=\omega_{i}+f(\{\theta\})+\sqrt{g(\{\theta\})}\xi_i(t)\,,
\end{equation}
where $\omega_{i}$ is the natural frequency for $i=1,2,...,N$ oscillators. The drift force $f(\{\theta\})$ and noise strength $g(\{\theta\})$ are general functions of phases $\{\theta\}=\theta_{1},...,\theta_{N}$, and $\xi_{i}$ is a Gaussian white noise that obeys the following relation:
\begin{equation}
\label{fdt}
\langle\xi_i(t)\xi_j(t')\rangle=2D\delta_{ij}\delta(t-t')\quad\mbox{with}\quad\langle\xi_i(t)\rangle=0\,,
\end{equation}
where $D$ is the dispersion of oscillators. The simplest dynamic equation for the phase oscillators was established by Kuramoto \cite{Kuramoto1}, with drift force given by
\begin{equation}
f(\{\theta\})={K \over N} \sum_{j=1}^N \sin(\theta_j - \theta_i)\,,
 \end{equation}
and $g(\{\theta\})=0$, where $K$ is the coupling strength. When $K>0$, the interaction is attractive. The signal of the force acting on {\it i}th oscillator is opposite to the displacement of this oscillator with respect to {\it j}th oscillator. For $K<0$, the interaction is repulsive. A simpler phase oscillator model with additive noise was proposed by Sakaguchi \cite{Sakaguchi}, where $g(\{\theta\})=1$.

We can usually define an order parameter $r=r(t)$ for the system, and its average phase $\psi=\psi(t)$ is given by
\begin{equation}
\label{mf}
re^{i\psi}=\frac{1}{N}\sum^{N}_{j=1}e^{i\theta_{j}}\,\,,
\end{equation}
where $r$ measures the phase coherence, {\it i.e.}, for $r=1$, the system is fully synchronized, whereas for $r=0$, the system is fully incoherent. A partially synchronized state is obtained when $0< r < 1$. In terms of Eq. (\ref{mf}), we can express $f(\{\theta\})=f(\theta)=rK\sin(\psi -\theta_{i})$, where the oscillators of the system interact with oscillator $\theta_i$. However, now their action is no longer considered individually but in terms of mean field quantities $r$ and $\psi$, which concern the state of all the phases. Note that when we express the interaction of the ensemble of oscillators with oscillator $\theta_ {i}$ in terms of mean field quantities, the index of the oscillator "i" can be omitted, since they now have phase oscillator $\theta$ interacting with average phase $\psi$ and with intensity modulated by order parameter $r$. Thus, we can express $g(\{\theta\})=g(\theta)$, assuming that function $g(\theta)$ can also be written in terms of mean field quantities where we can neglect the action of individual oscillators on the phase of oscillator $\theta_i$. Now we adopt an identical natural frequency $\omega_{i}=\omega$, which allows the system to reach the thermodynamic equilibrium \cite{Gupta,Bonilla2}. On the rotating frame, we can set $\omega=0$. It follows that Eq. (\ref{eq:kura_model}) can be rewritten as
\begin{equation}
\dot{\theta}=f(\theta)+\sqrt{g(\theta)}\xi(t)\,.
\label{eq:lang_mult}
\end{equation}

Our objective is to study the thermodynamics of phase oscillators starting from the general Ito phase equation to an oscillator system with strong limit cycle attractor \cite{Lai,Yoshimura,Teramae2} in which Eq. (\ref{eq:lang_mult}) is given by
\begin{equation}
\label{eq:lang_mult2}
\dot{\theta}= DZ(\theta)Z'(\theta)+Z(\theta)\xi(t)\,,
\end{equation}
where $Z'(\theta)=\partial Z(\theta)/\partial\theta$ and $D$ is the diffusion such that functions $f(\theta)$ and $g(\theta)$ result in
\begin{eqnarray}
&&f(\theta)=DZ(\theta)Z'(\theta)=rK\sin(\psi -\theta)\,,\\
\nonumber\\
&&\sqrt{g(\theta)}=Z(\theta)=\sqrt{1+r\sigma\cos(\psi - \theta)}\quad\text{as}\quad-1\leq\sigma\leq 1\,,
\end{eqnarray}
from which is found the relation $\sigma D=2K$, with $D>0$. Parameter $\sigma$ is the noise coupling that determines the intensity of $\sqrt{g(\theta)}$, i.e., the global modulation of the multiplicative noise. Note that for $\sigma\neq 0$, the system's noise intensity depends on phase $\theta$ of the oscillators. For $\sigma=0$ and $g(\theta)=1$, we retrieve the conventional Sakaguchi model with additive noise.

\section{Thermodynamic equilibrium and phase density}

In order to study the thermodynamics of the model, we write in Ito prescription the corresponding Fokker--Planck equation from Eq. (\ref{eq:lang_mult}) in the form
\begin{equation}
\frac{\partial\rho}{\partial t} = D\frac{\partial^2}{\partial\theta^2}[g(\theta)\rho]-\frac{\partial}{\partial\theta}[f(\theta)\rho]\,.
\label{eq:F-P_mult}
\end{equation}

To demonstrate that the system precisely satisfies the thermodynamic equilibrium condition, it is more convenient to transform the Langevin equation with multiplicative noise to an equation with additive noise and interpret the dynamics of the system as diffusion under the action potential. In this case, we can always do this for a one-dimensional system, as given by Eq. (\ref{eq:lang_mult}) and time-independent functions $f$ and $g$ \cite{Risken}. Thus, let us make this transformation in the Fokker--Planck Eq. (\ref{eq:F-P_mult}) using the following change of variables
\begin{equation}
\phi(\theta)=\frac{1}{\sqrt{D}}\int_0^\theta\frac{d\theta'}{\sqrt{g(\theta')}}\,
\label{eq:transf_variaveis}
\end{equation}
such that the new equation that governs the temporal evolution of distribution $P(\phi,t)$ for new variables can be written as
\begin{equation}
\frac{\partial P(\phi,t)}{\partial t}=\frac{\partial^2 P}{\partial\phi^2}-\frac{\partial}{\partial\phi}[{\cal F}(\phi)P]\,,
\label{eq:F-P_nova}
\end{equation}
where
\begin{equation}				
{\cal F}(\phi)=\frac{1}{\sqrt{D}}\left[\frac{f(\theta)}{\sqrt{g(\theta)}}
-D\frac{\partial}{\partial\theta}\sqrt{g(\theta)}\right]_{\theta=\theta(\phi)}\,.
\label{eq:drift_transf}
\end{equation}
Therefore, the corresponding Langevin Eq. (\ref{eq:F-P_nova}) is given by
\begin{equation}
\label{lang2}
\dot{\phi}={\cal F}(\phi)+\xi(t)\,,
\end{equation}
where $\xi$ is a Gaussian additive noise with zero average and unit variance. As the diffusion term is now a constant, the condition for Eq. (\ref{lang2}) to obey the detailed balance is that drift term ${\cal F}(\phi)$ should satisfy the potential function condition. For obtaining an analytical expression for ${\cal F}(\phi)$, it is necessary to evaluate the integral Eq. (\ref{eq:transf_variaveis}) and reverse the resulting expression for $\theta(\phi)$, which appears in Eq. (\ref{eq:drift_transf}). However, we can determine whether drift term ${\cal F}(\phi)$ is conservative and infer if the system reaches thermal equilibrium even without an explicit expression for this function \cite{Russel}. First, we note that if ${\cal F}(\phi)$ is conservative, the work in a closed path should be null
\begin{equation}
\oint {\cal F}(\phi)d\phi = 0\,,
\end{equation}
where we can write the ansatz to ${\cal F}$ as
\begin{equation}
{\cal F}(\phi)=\lambda-\frac{\partial}{\partial\phi}V(\phi)\,,
\label{eq:force}
\end{equation}
such that the first term on the right side is a force of not constant balance and the second term is a conservative force derived from potential $V(\phi)$. To formally show that $\lambda = 0$, let us write ${\cal F}(\phi)$ as a Fourier series
\begin{equation}				
{\cal F}(\phi)=\frac{1}{2}a_0+\sum_{n=1}^\infty\left[a_n\cos\left(\frac{n\phi}{\mathcal{T}}\right)
+b_n\sin\left(\frac{n\phi}{\mathcal{T}}\right)\right]\,,
\label{eq:fourier_force}
\end{equation}
where using Eq. (\ref{eq:transf_variaveis})
\begin{equation}
\mathcal{T}=\phi(\pi)-\phi(0)=\frac{1}{2\sqrt{D}}\int_{-\pi}^\pi \frac{d\theta'}{\sqrt{g(\theta')}}\,
\end{equation}
is half the transformed range $[-\pi,\pi]$. The coefficients $a_n$ e $b_n$ are commonly given by
\begin{eqnarray}
a_n=\frac{1}{\mathcal{T}}\int_{-\mathcal{T}}^\mathcal{T}{\cal F}(\phi)\cos\left(\frac{n\phi}{\mathcal{T}}\right)d\phi \nonumber\\
b_n=\frac{1}{\mathcal{T}}\int_{-\mathcal{T}}^\mathcal{T}{\cal F}(\phi)\sin\left(\frac{n\phi}{\mathcal{T}}\right)d\phi\,.
\end{eqnarray}
Thus, comparing Eq. (\ref{eq:force}) with the Fourier series given by Eq. (\ref{eq:fourier_force}), we see that $\lambda=a_0/2$ so that the question of determining whether ${\cal F}(\phi)$ is conservative and whether the system reaches thermal equilibrium is equivalent to finding the condition $a_0=0$. Then taking
\begin{equation}
a_0=\frac{1}{\mathcal{T}}\int_{-\mathcal{T}}^\mathcal{T}{\cal F}(\phi)d\phi\,
\end{equation}
and applying the change of variables, we get
\begin{equation}
a_0=\frac{1}{\mathcal{T}}\int_{\phi(-\mathcal{T})}^{\phi(\mathcal{T})}{\cal F}(\phi(\theta))\frac{d\phi}{d\theta}d\theta\,.
\end{equation}
Then by defining $\theta(\pm\mathcal{T})=\pm\pi$ and from ${\cal F}(\phi)$ from Eq. (\ref{eq:drift_transf}), we can write coefficient $a_0$ as
\begin{equation}
a_0 = \frac{1}{\mathcal{T}D}\int_{-\pi}^\pi\left[\frac{f(\theta)}{g(\theta)}-\frac{D}{2}\frac{\partial}{\partial\theta}\ln g(\theta)\right]d\theta\,.
\end{equation}
Note that the second term of the integrand vanishes due to periodicity of $g(\theta)$. The integral of the first term is explicitly given by
\begin{equation}
\int_{-\pi}^\pi\frac{f(\theta)}{g(\theta)}d\theta = \int_{-\pi}^\pi\frac{rK\sin(\psi-\theta)}{1+r\sigma\cos(\psi-\theta)}d\theta\,.
\end{equation}
Thus, by performing a simple integration, we find
\begin{equation}
a_0=\frac{K}{\mathcal{T}D\sigma}\left[\ln(1+r\sigma\cos(\psi-\theta)\right]_{-\pi}^{\pi}=0\,.
\end{equation}
Therefore, it demonstrates that on the system described by Eq. (\ref{eq:lang_mult}) act conservative forces ${\cal F}(\phi)=-dV/d\phi$, which relax to a state of thermodynamic equilibrium at $t\rightarrow\infty$. Note that the stationary distribution density $\rho(\theta,\infty)=\rho_{s}(\theta)$ satisfies this requirement. Indeed, taking the Fokker--Planck Eq. (\ref{eq:F-P_mult}) in the continuous limit $N\rightarrow\infty$, we have
\begin{equation}
\label{fokkerg}
{\partial\rho\over\partial t} =D{\partial^2\over\partial\theta^2}\big[(1+r\sigma\cos(\psi-\theta))\rho\big]-{\partial\over\partial\theta}\big[rK\sin(\psi-\theta)\rho\big]\,.
\end{equation}
This equation has an exact analytical expression for the stationary distribution $\rho_{s}(\theta)$, given by
\begin{equation}
\label{hypermises}
\rho_{s}(\theta)={\cal N}^{-1}\big[z+\sgn(\sigma)\sqrt{z^2-1}\cos(\psi-\theta)\big]^\nu\,,
\end{equation}
where $\sgn(\sigma)$ is the sign function. The normalization constant ${\cal N}$ and parameters $z$ and $\nu$ are
\begin{eqnarray}
\label{constn}
        &&{\cal N}=2\pi P_{\nu}^{0}(z)\,,\\
        &&z=(1-\sigma^{2}r^2)^{-1/2}\,,\\
        &&\nu=\frac{K}{D\sigma}-1\,,
 \end{eqnarray}
where $P_{\nu}^{0}(z)$ is the associated Legendre function of zero order. As discussed, stationary distribution $\rho_{s}(\theta)$ obeys the thermodynamic equilibrium condition. Distributions of power law as $\rho_{s}(\theta)$ have been currently found in many complex systems \cite{Hongler} and can be regarded as a more general case of the exponential behavior of the Boltzmann--Gibbs distributions. These distributions can be characterized as both the stationary states of non-equilibrium and thermodynamic equilibrium \cite{Ran}. We demonstrate that $\rho_{s}$ obeys the thermodynamic equilibrium, {\it i.e.}, it results in a null probability current density $J_{s}=0$ in the configuration space. See this demonstration in Appendix A.

\section{Order parameter and temperature}
We can now obtain the general expression for order parameter $r$, which results in
\begin{equation}
\label{order2}
r=\int_{0}^{2\pi}e^{i(\theta-\psi)}\rho_{s}(\theta)d\theta=\frac{\sgn{(\sigma)}}{1+\nu}\frac{P_{\nu}^{1}(z)}{P_{\nu}^{0}(z)}\,.
\end{equation}
This allows us to obtain critical coupling $K_{c}$ of the oscillator system. Hence, by taking Eq. (\ref{order2}) in the critical region $r\approx\nu r\sigma/2$ and using $\nu=K_{c}/D\sigma-1$, it follows that critical coupling $K_{c}$ is
\begin{equation}
\label{criticalK}
K_{c}=D(\sigma+2)\,.
\end{equation}
See Appendix B for more details. Indeed, Eq. (\ref{criticalK}) shows that for $\sigma=0$, we retrieve the classic $K_{c}=2D$ from the Kuramoto model with additive noise. For $\sigma<0$, noise coupling tends to weaken the dispersive action on the oscillators, reducing critical coupling $K_{c}$. On the other hand, it has an inverse effect for $\sigma> 0$.

The effective temperature $T_{eff}$ of the system is now defined as
\begin{equation}
\label{temp1}
T_{eff}\equiv\frac{K_{c}}{K}=(\frac{\sigma}{2}+1)T\,,
\end{equation}
where $T=\frac{2D}{K}$ is the temperature for $\sigma=0$, {\it i.e.}, for the model with additive noise. Therefore, for any $\sigma$, the critical effective temperature is $T_{effc}=1$. It is also useful to redefine parameter $\nu$ as
\begin{equation}
\label{nudef}
\nu=\frac{2}{T\sigma}-1\,.
\end{equation}

\begin{figure}[!h]
\label{fig1}
\centering\rotatebox{-90}{\resizebox{6.5cm}{!}{\includegraphics{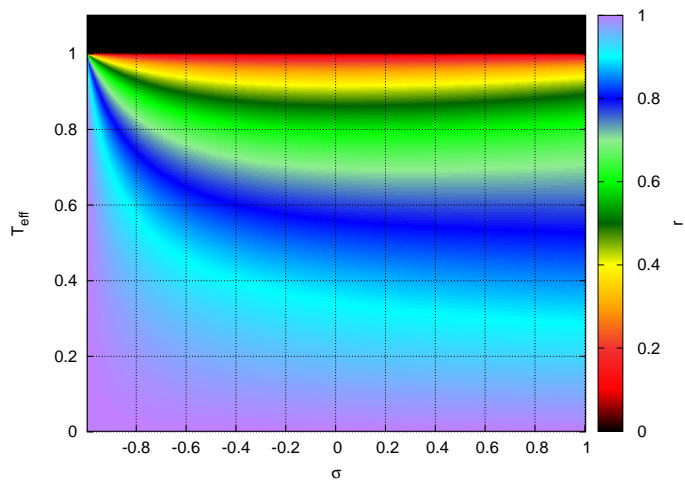}}}
\caption{Spectrum of order parameter $r$ in space $(T_{eff},\sigma)$. The black region above $T_{eff}=T_{effc}=1$ defines the region without order $r=0$. Note that the region $T_{effc}<1$ contains the entire synchronized region including that for which $T>1$; we call this as a parasynchronized phase.}
\end{figure}

Figure 1 shows the spectrum of order parameter $r$ in space $(\sigma,T_{eff})$. It has been implemented numerically with self-consistent calculation of Eq. (\ref{order2}), also using the results of Appendix B. The black strip corresponds to the region $r=0$, while the far violet is the region with the maximum order that corresponds to $r \approx 1$. This clearly shows the asymmetry between the regions with $\sigma >0$ and $\sigma < 0$. Moreover, note that negative values of $\sigma$ favor synchronization of the oscillator system. It is very instructive to observe that $\sigma$, as defined in Eq. (\ref{temp1}), induces the existence of a parasynchronized phase in the system, {\it i.e.}, it leads to the existence of order for $T>1$ up to $T_{eff} <1$, which occurs for values of $\sigma$ in the region between the curves $T_{eff}=\sigma/2+1$ and $T_{eff}=1$. Indeed, the properties of the parasynchronized phase in the system will become clearer in Figure 2 below.

\begin{figure}[!h]
\label{fig2}
\centering\rotatebox{-90}{\resizebox{6.5cm}{!}{\includegraphics{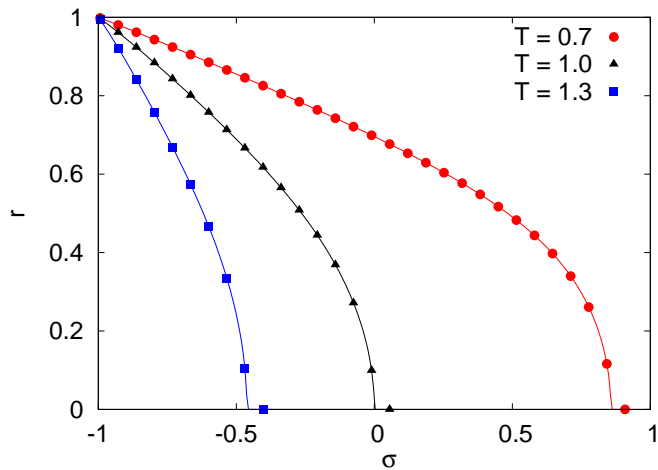}}}
\caption{\label{fig2} Order parameter $r$ as a function of noise coupling $\sigma$ for isotherms $T=1$, $T>1$, and $T<1$. The region $T > 1$ characterizes the parasynchronized phase with $\sigma<0$, and the region $T\leq 1$ corresponds to the synchronized phase.}
\end{figure}

Figure 2 shows the behavior of $r$ as a function of coupling $\sigma$ for the isotherms. For all curves, $T_{eff}<1$. We see that for fixed temperature $T$, $r$ is a decreasing function of $\sigma$. Here we observe that the curve $T=1$ separates the aforementioned internal region referred to as a parasynchronized phase, established for $T > 1$, from the synchronized phase with $T \leq 1$. The parasynchronized phase exists only for $\sigma < 0$ and for all those curves, $T_{eff} \leq 1$ is verified. The asymmetry in the synchronization behavior for the values of $\sigma$ and $-\sigma$ is clearly shown, as discussed in relation to Figure 1. It is also important to note that there is a second-order phase transition induced solely by the effects of the multiplicative noise.

\begin{figure}[!h]
\label{fig3}
\centering\rotatebox{-90}{\resizebox{6.5cm}{!}{\includegraphics{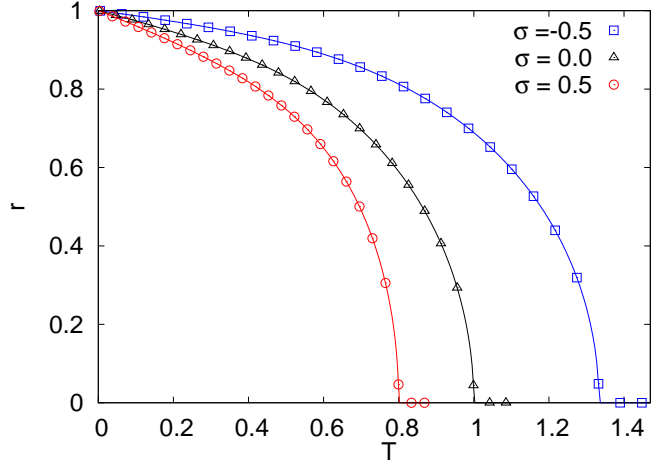}}}
\caption{\label{fig3} Order parameter $r$ as a function of temperature $T$ for values of $\sigma$. Note that all curves display a typical mean field behavior.}
\end{figure}

Figure 3 shows the behavior of order parameter $r$ as a function of temperature $T$. From left to right, we have $\sigma=+0.5$, $0$, and $-0.5$. It shows the typical behavior where the order parameter decreases with temperature. For $\sigma=0$, the middle curve corresponds to simple additive noise for which $T=T_{eff}$ and consequently critical temperatures $T_c$ and $T_{effc}$ are the same. The curves with non-null noise coupling present asymmetry between $\sigma=0.5$ (with $T_c=4/5$) and $\sigma=-0.5$ (with $T_c=4/3$), the latter presents order even for $T>1$. Note that from Eq. (\ref{temp1}), the two curves respectively have $T_{effc}=(1+0.5/2)4/5=1$ and $T_{effc}=(1-0.5/2)4/3=1$. It also becomes clear that the concept of effective temperature is useful to scale all these curves to a unique transition point $T_{effc}=1$.

Once we identify noise coupling $\sigma$ as being responsible for inducing a continuous phase transition in the oscillator system, we can think about the existence of a thermodynamic field associated to $\sigma$ able to induce synchronization. Thus, by considering the internal energy of the system as a function of entropy and order parameter $U=U(S,r)$, we can write the following equation:
\begin{equation}
dU=TdS-H_{s}dr\,,
\end{equation}
where $H_{s}$ is a new quantity, which we refer to as a synchronization field associated with the action of the multiplicative noise on the system. Note that the negative sign comes from the fact that an increase in the order parameter, by maintaining constant entropy, leads to a decrease in the internal energy of the system, i.e., by increasing $r$, we should increase the order of the system.

The Helmholtz free energy $F=F(T,r)$ is obtained by the Legendre transform of the internal energy, which results in the equation
\begin{equation}
\label{Fhelm}
dF=-SdT-H_{s}dr\,.
\end{equation}
This is the first law of thermodynamics for phase synchronization with internal multiplicative noise. Thus, based on the free energy, we can obtain entropy $S$ and synchronization field $H_{s}$ in the following sections.

\section{Entropy and free energy}

We begin by determining entropy directly from Gibbs' definition
\begin{eqnarray}
S&=&-\int\rho_{s}(\theta)\ln\rho_{s}(\theta)d\theta\nonumber\\
&=&\ln {\cal N}-\nu {\cal N}^{-1}\int_{0}^{2\pi}\big[\lambda(z,\theta)\big]^{\nu}\ln[\lambda(z,\theta)\big]d\theta\nonumber\\
&=&\ln {\cal N}-\nu {\cal N}^{-1}\lim_{\varphi\rightarrow0}{\partial\over\partial\varphi}\int_{0}^{2\pi}[\lambda(z,\theta)\big]^{\nu+\varphi}d\theta\nonumber\\
&=&\ln {\cal N}-2\pi\nu {\cal N}^{-1}\lim_{\varphi\rightarrow 0}\frac{\partial }{\partial\varphi}P^{0}_{\nu+\varphi}(z)\nonumber\\
&=&\Big(1-\nu\frac{\partial}{\partial\nu}\Big)\ln\big[2\pi P_{\nu}^{0}(z)\big]\,,
\label{S1}
\end{eqnarray}
where $\rho_{s}$ is the stationary distribution Eq. (\ref{hypermises}). We assume the Boltzmann constant $k_{B}=1$. Here $\lambda(z,\theta)=z+\sgn(\sigma)\sqrt{z^2-1}\cos(\psi-\theta)$ and $P_{\nu}^{0}(z)$ are the associated Legendre functions of zero order. We use the normalization condition ${\cal N}^{-1}\int_{0}^{2\pi}\rho(\theta)d\theta=1$ as well as the relation $\lim_{\varphi\rightarrow 0}\frac{\partial }{\partial\varphi}P^{0}_{\nu+\varphi}(z)=\Big[\frac{\partial }{\partial\varphi}P^{0}_{\varphi}(z)\Big]_{\varphi=\nu}$, see Cohl \cite{Cohl}.

The free energy $F$ of the system can be obtained from the equilibrium statistical mechanics by the expression
\begin{equation}
F=-T\ln Z = -T{\ln\cal N}=-T\ln[2\pi P_{\nu}^{0}(z)]\,,
\label{F_geral}
\end{equation}
where partition function $Z$ is equivalent to the normalization constant $\cal N$ according to distribution Eq. (\ref{hypermises}). We can now directly derive the entropy Eq. (\ref{S1}) of free energy $F$ employing Eq. (\ref{Fhelm}) as
\begin{equation}
S=-\left(\frac{\partial F}{\partial T}\right)_r = \ln[2\pi P_\nu^0(z)]+T\frac{\partial}{\partial T}\ln[2\pi P_\nu^0(z)]\,,
\end{equation}
where we can express the above equation in terms of parameter $\nu$. By taking the transformation in the derived $\partial/\partial T = \partial\nu/\partial T\left(\partial/\partial\nu\right)$ and using Eq. (\ref{nudef}), we have $\partial\nu/\partial T = -(\nu+1)/T$, which results in
\begin{equation}
S=\ln[2\pi P_\nu^0(z)]-(\nu+1)\frac{\partial}{\partial \nu}\ln[2\pi P_\nu^0(z)].
\end{equation}
Hence, performing the change of variables $\nu'=-(\nu+1)$, the derivative with respect to $\nu$ changes as $\partial/\partial\nu = \partial\nu'/\partial\nu\left(\partial/\partial\nu'\right)=-\partial/\partial\nu'$ such that
\begin{equation}
S=\ln[2\pi P_{-\nu'-1}^0(z)]-\nu'\frac{\partial}{\partial \nu'}\ln[2\pi P_{-\nu'-1}^0(z)].
\end{equation}
Using the property of Legendre functions $P_{-\nu-1}^{\pm m}(z)=P_\nu^{\pm m}(z)$ and changing $\nu'\rightarrow\nu$, we find entropy expression
\begin{equation}
\label{hyperentropy}
S=\left(1-\nu\frac{\partial}{\partial\nu} \right)\ln[2\pi P_\nu^0(z)]\,,
\end{equation}
which is identical to Eq. (\ref{S1}). In this sense, the free energy Eq. (\ref{F_geral}) and entropy Eq. (\ref{hyperentropy}) denote the general thermodynamic equilibrium properties of the system as well as affords us the opportunity to study the thermodynamic properties of synchronization for systems far beyond conventional order parameter analysis.

From free energy and entropy, we can also derive the internal energy of the system:
\begin{equation}
\label{internalener}
U=-T\nu\frac{\partial}{\partial\nu}\ln[2\pi P_\nu^0(z)]\,,
\end{equation}
where we have used the thermodynamic equation $F=U-TS$.
\begin{figure}[!h]
\label{fig4}
\centering\rotatebox{-90}{\resizebox{6.5cm}{!}{\includegraphics{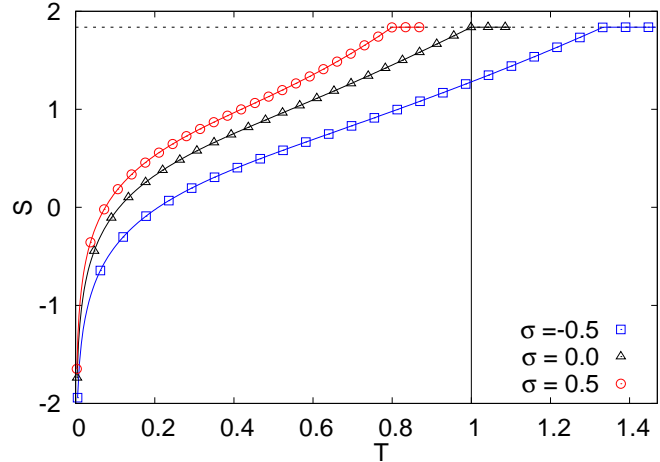}}}
\caption{\label{fig4} Entropy as a function of temperature $T$ for values of $\sigma$. Note that all entropy reach the maximum value at $\ln(2\pi)$, which is the full incoherent state. All entropy becomes $S=-\infty$ at $T=0$, which is the full state synchronizing.}
\end{figure}

Figure 4 shows the behavior of entropy Eq. (\ref{hyperentropy}) as a function of temperature $T$. In the critical effective temperature $T_{effc}=(\sigma/2+1)T=1$, all entropy will reach the maximum value at
\begin{equation}
\label{smax}
S_{max}= \ln{(2\pi)}\,,
\end{equation}
which corresponds to the value of the fully desynchronized state. It is important to note that for negative $\sigma$, there is a reduction in the value of critical coupling when compared to the coupling of the simple model with additive noise $K_{c}=2D$, implying that the system is more easily driven toward synchronization.

\subsection{Limits of Entropy }

The physical limits of Eq. (\ref{hyperentropy}) can be now analyzed. First, we consider the weak noise coupling condition $\sigma\approx 0$, which implies $\nu\rightarrow\infty$. In this case, entropy is given by
\begin{equation}
\label{entropyf}
S=\frac{1}{2}\ln\left({4\pi^2\cosh^{-1}(z)\over\sqrt{z^2-1}}\right)+\ln[I_{0}(\varepsilon)]-\varepsilon{I_{1}(\varepsilon)\over I_{0}(\varepsilon)}\,,
\end{equation}
where we use
\begin{equation}
\label{asymp}
P_{\nu}^{0}(z)\approx\Big({\cosh^{-1}(z)\over\sqrt{z^2-1}}\Big)^{1/2}I_{0}(\nu\cosh^{-1}(z)) \quad\mbox{as}\quad\nu\rightarrow\infty\,\,,
\end{equation}
in which $I_{n}(x)$ is the modified Bessel function of first kind of order $n$ and defined as
\begin{equation}
I_n(x)=\frac{1}{2\pi}\int_0^{2\pi}e^{x\cos(\theta)}\cos(n\theta)d\theta\,.
\end{equation}
Here for the asymptotic limit $\nu\rightarrow\infty$, the parameters in Eq. (\ref{entropyf}) are given by
\begin{eqnarray}
\label{defin21}
        &&\varepsilon\equiv\nu\cosh^{-1}(z)\,,\\
        &&z=(1-\sigma^2{r(\varepsilon)}^{2})^{-1/2}\,,\\
        &&r(\varepsilon)=\lim_{\nu\rightarrow\infty}r=\frac{I_{1}(\varepsilon)}{I_{0}(\varepsilon)}\,.
 \end{eqnarray}
The condition $\sigma=0$ is a particular case of Eq. (\ref{entropyf}), which is just the entropy for the Kuramoto model with additive noise
\begin{equation}
S=\ln\left[2\pi I_{0}(k)\right]-k\frac{I_{1}(k)}{I_{0}(k)},
\end{equation}
with
\begin{eqnarray}
\label{defin31}
        &&k=\lim_{\sigma\rightarrow 0}\varepsilon=\frac{2r(k)}{T}\,,\\
        &&r(k)=\lim_{\sigma\rightarrow 0}r(\varepsilon)=\frac{I_{1}(k)}{I_{0}(k)}\,.\\
 \end{eqnarray}
The condition $T\approx 0$ implies $\nu\rightarrow\infty$, which is also a particular case of Eq. (\ref{entropyf}). Thus, taking $T=0$ in Eq. (\ref{entropyf}), we obtain
\begin{equation}
S(T=0)=\lim_{\varepsilon\rightarrow\infty}S\rightarrow-\infty\,.
\end{equation}
Note that $T=0$ implies $\varepsilon\rightarrow\infty$, where we have used $I_{0}(\varepsilon)\sim e^{\varepsilon}/\sqrt{2\pi\varepsilon}$ in Eq. (\ref{entropyf}). At this limit, entropy assumes the same behavior as the entropy of a classical ideal gas at $T=0$. We also see that for $T=0$, the phases of all oscillators have the same value and consequently the phase density of the system is equivalent to a Dirac delta function. Indeed, this result is expected since the oscillator system treated here is described in classical phase space.

\section{Specific heat}

Having obtained free energy and analyzed entropy function in the above section, we can study specific heat of the system by keeping fixed noise coupling $\sigma$. This requires determining
\begin{eqnarray}
C_{\sigma}&=&T\left(\frac{\partial S}{\partial T}\right)_{\sigma}\nonumber\\
\nonumber\\
&=&T\left[\left(\frac{\partial S}{\partial\nu}\right)_{z}\frac{\partial\nu}{\partial T}+\left(\frac{\partial S}{\partial z}\right)_{\nu}\frac{\partial z}{\partial T}\right]\nonumber\\
\nonumber\\
&=&T\left[\left(\frac{\partial S}{\partial\nu}\right)_{z}+\frac{1}{\left[1-\left(\frac{\partial z}{\partial z}\right)_{\nu}\right]}\left(\frac{\partial z}{\partial\nu}\right)_{z}\left(\frac{\partial S}{\partial z}\right)_{\nu}\right]\frac{\partial \nu}{\partial T}\,.
\end{eqnarray}
Here we assume $k_{B}=1$. Note that $z$ is a transcendental function, $z\rightarrow z(\nu,z)$. Then we can rewrite $\partial z/\partial T$ as
\begin{equation}
\frac{\partial z}{\partial T}=\left(\frac{\partial z}{\partial\nu}\right)_{z}\frac{\partial \nu}{\partial T}+\left(\frac{\partial z}{\partial z}\right)_{\nu}\frac{\partial z}{\partial T}=\frac{1}{\left[1-\left(\frac{\partial z}{\partial z}\right)_{\nu}\right]}\left(\frac{\partial z}{\partial\nu}\right)_{z}\frac{\partial \nu}{\partial T}.
\end{equation}
The partial derivatives are
\begin{eqnarray}
&&\left(\frac{\partial S}{\partial\nu}\right)_{z}=\frac{\sgn(\sigma)}{\sqrt{z^2-1}}\left[r-\nu(\nu+1)\frac{\partial r}{\partial\nu}\right]\,\,,\\
\nonumber\\
&&\left(\frac{\partial S}{\partial z}\right)_{\nu}=-\nu\frac{\partial^{2}}{\partial\nu^{2}}\ln\left[2\pi P^{0}_{\nu}(z)\right]\,\,,\\
\nonumber\\
&&\left(\frac{\partial z}{\partial z}\right)_{\nu}=-\varsigma(z)\frac{\partial r}{\partial z}\,\,,\\
\nonumber\\
&&\left(\frac{\partial z}{\partial\nu}\right)_{z}=-\varsigma(z)\frac{\partial r}{\partial\nu}\,\,,
\end{eqnarray}
where we have used the Eqs. (\ref{order2}), (\ref{hyperentropy}), and
\begin{eqnarray}
&&z=\cosh\left[\tanh^{-1}\left(\frac{\sgn{(\sigma)}}{1+\nu}\frac{P_{\nu}^{1}(z)}{P_{\nu}^{0}(z)}\right)\right]\nonumber\\
&&\varsigma(z)=z\sigma Q^{1}_{0}(r\sigma)\sinh\big[Q_{0}^{0}(r\sigma)\big]\,,
\end{eqnarray}
in which $Q_{0}^{0}(x)$ and $Q^{1}_{0}(x)$ are the associated Legendre functions of the second kind~\cite{Erdelyi}. Hence, specific heat is given by
\begin{eqnarray}
\label{thermalc0}
\!\!\!\!\!\!\!\!C_{\sigma}=\frac{\sgn(\sigma)\nu(\nu+1)\varsigma(z)}{\sqrt{z^2-1}\left(1+\varsigma(z)\frac{\partial r}{\partial z}\right)}\left[r\frac{\partial r}{\partial\nu}-\nu(\nu+1)\left(\frac{\partial r}{\partial\nu}\right)^2\right]+\nu(\nu+1)\frac{\partial^2}{\partial\nu^{2}}\ln\left[2\pi P^{0}_{\nu}(z)\right].
\end{eqnarray}

Figure 5 shows specific heat Eq. (\ref{thermalc0}) as a function of temperature $T$ for values of $\sigma$. We see that all curves start at $C_{\sigma}( T \rightarrow 0)=\frac{1}{2}$ in precise accordance with our analytic result, see Eq. (\ref{C0}). We observe that specific heat grows continuously until it reaches a maximum value at the critical effective temperature $T_{effc}=(\sigma/2+1)T=1$. The increase in entropy required to reduce synchronization results in $C_{\sigma}(T>0)>C_{\sigma}(0)$. As expected, note that for $T>T_{effc}$, the system achieves a totally desynchronized state and specific heat is therefore reduced to zero. Note also that thermodynamic stability $C_{\sigma}>0$ is immediately satisfied for all $\sigma$.

\begin{figure}[!h]
\label{fig5}
\centering\rotatebox{-90}{\resizebox{6.5cm}{!}{\includegraphics{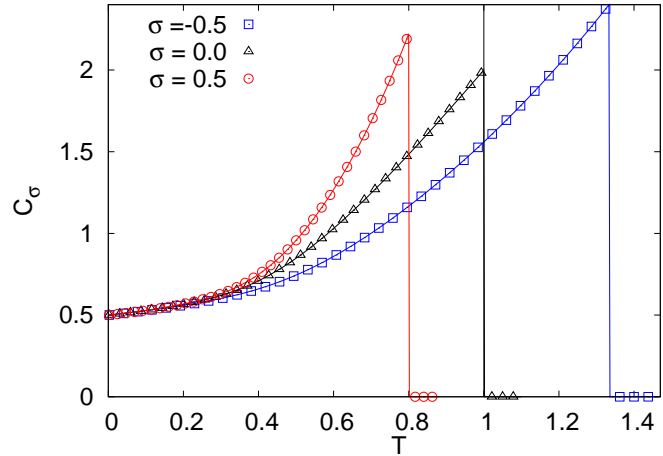}}}
\caption{\label{fig5} Specific heat $C_{\sigma}(T)$ as a function of temperature $T$ for values of $\sigma$. All curves start at $C_{\sigma}=\frac{1}{2}$ for $T=0$ and grow until it reaches the maximum value at $T_{effc}=1$. For $T>T_{effc}$, $C_{\sigma}=0$, as expected.}
\end{figure}

\subsection{Limits of specific heat}

From entropy Eq. (\ref{entropyf}), we derive specific heat for $\sigma\approx 0$, which is given by
\begin{equation}
\label{thermalf}
C_{\sigma}=\frac{\sigma\varepsilon}{2}\left[\frac{2\varepsilon}{\sigma z^2}-\frac{1}{\cosh^{-1}(z)}+\frac{z}{\sqrt{z^2-1}}\right]\frac{\Phi(\varepsilon)}{1-\nu\sigma\Phi(\varepsilon)}\,,
\end{equation}
where
\begin{equation}
\Phi(\varepsilon)=\frac{1}{\sigma}\frac{\partial}{\partial\varepsilon}\cosh^{-1}(z(\varepsilon)).
\end{equation}
In particular for $\sigma=0$, Eq. (\ref{thermalf}) yields
\begin{equation}
\label{Cadd}
C_{\sigma=0}=\lim_{\sigma\rightarrow 0}C_{\sigma}=\frac{k^{2}\Phi(k)}{1-\frac{2}{T}\Phi(k)}\,.
\end{equation}
where we have
\begin{equation}
\Phi(k)=\lim_{\sigma\rightarrow 0}\Phi(\varepsilon)=\frac{\partial r(k)}{\partial k}.
\end{equation}
We call attention to the fact that Eq. (\ref{Cadd}) is the specific heat for the model with only additive noise. Finally, the specific heat for $T=0$ is obtained from Eq. (\ref{thermalf}), which results in
\begin{equation}
\label{C0}
C_{\sigma}(T=0)=\lim_{\varepsilon\rightarrow\infty}C_{\sigma}=\frac{1}{2}.
\end{equation}
This is precisely what we obtain for the numerical calculations, see Figure 5. Note that this is the same value for specific heat at constant volume $C_{v}$ for a classical ideal gas.

\section{Synchronization field and susceptibility}

We can now define the synchronization field $H_{s}$ from thermodynamics Eq. (\ref{Fhelm}) as
\begin{equation}
\label{fullH}
H_{s}=-\left(\frac{\partial F}{\partial r}\right)_{T}\,,
\end{equation}
in which free energy $F$ is given by Eq. (\ref{F_geral}). In Eq. (\ref{fullH}), an increase in order parameter $r$ results in a decrease in free energy $F$, where $\partial F/\partial r<0$, and in this condition, we have $H_{s}>0$. Note that the synchronization field should not be simply understood as a conventional external field acting on the oscillator system. This is analogous to pressure behavior in classical thermodynamic systems $p=-\left(\partial F/\partial V\right)_{T}$, in which volume $V$ is analogous to order parameter $r$. Nevertheless, we call attention to the complexity of this oscillator system, where for low temperature $T \rightarrow 0$, entropy behaves as $S(T) \rightarrow \ln(T)\rightarrow-\infty$ and specific heat as $C_{\sigma}\rightarrow1/2$, which bears more similarity to a classical gas, while for higher temperatures is found a similarity with the magnetic system and complex liquids.

To better understand these properties, we first need to determine the full expression for $H_s$ requiring that $z=(1-\sigma^2r^2)^{-1/2}$, so the derivative with respect to $r$ transforms as $\partial/\partial r=z^3\sigma^2r\left(\partial/\partial z\right)$. Thus, using Eq. (\ref{fullH}), we obtain the expression for the synchronization field as
\begin{equation}
H_s=Tz^3\sigma^2r\frac{\partial}{\partial z}\ln[2\pi P_\nu^0(z)]
=\frac{T(1+\nu)z^3\sigma^2r^{2}\sgn(\sigma)}{\sqrt{z^2-1}}\label{fullH2}\,.
\end{equation}
\vspace{0.5cm}
Here we use Eq. (\ref{order2}) and the relationship $P_\nu^1(z)=(z^2-1)^{1/2}dP_{\nu}^{0}(z)/dz$.

It is now important to establish the dependence of field $H_{s}$ on parameter $\sigma$, i.e., how field $H_{s}$ is associated with the noise effect of the system. To accomplish this, we need to decompose $H_{s}$ as
\begin{equation}
\label{Hbroken}
H_{s}= H_{0} + H_{\sigma}\,,
\end{equation}
where we define the internal synchronization field $H_{0}$ as
\begin{equation}
\label{Hmf}
H_{0}=H_{s}\left(\sigma=0,r,T\right)\,.
\end{equation}
This corresponds to part of field $H_{s}$, which does not depend explicitly on $\sigma$, i.e, it is intrinsically associated with the Gaussian white noise behavior. Thus, the expression of $H_{0}$ is obtained by taking the limit $\sigma\rightarrow 0$ in Eq. (\ref{fullH2}), which results in
\begin{equation}
\label{HGmf}
H_{0}=\lim_{\sigma\rightarrow 0}H_{s}=2r=2\frac{I_{1}(2r/T)}{I_{0}(2r/T)}\,,
\end{equation}
which takes in account only the effect of additive noise on the system.

We define the external synchronization field $H_{\sigma}$ as
\begin{equation}
\label{Hsigma}
H_{\sigma}=H_{s}(\sigma,r,T)-H_{0}\,.
\end{equation}
This corresponds to part of field $H_{s}$, which explicitly depends on $\sigma$. Indeed, for $\sigma\neq 0$, the synchronization state of the system is a function of the phase of each oscillator, as established by the multiplicative noise. It means field $H_{\sigma}$ can be interpreted as a thermodynamic field that is directly associated to the phase-dependence effect of the noise on synchronization.

Here we can make an interesting analogy between the conventional Ising model in magnetism and the decomposition of synchronization field $H_ {s}$, as established by the Eq. (\ref{Hbroken}). The mean field as usually defined for an Ising system is an effective field given by $H_{eff}=c_0m +H'$, where $c_0$ is a constant, $m$ denotes magnetization, and $H'$ is a typical external field that does not depend on the internal parameters.
Note that we have defined here a synchronization field $H_{s}=2r+H_{\sigma}$ that plays the role of an effective field for which the internal field $H_{0}=2r$ is analogous to the $c_0m$ while the external field $H_{\sigma}$ is similar to $H'$. Nevertheless, the external field $H_{\sigma}$ is far from being a constant, it is a function that depends intrinsically on $r$, $T$, and $\sigma$.

\subsection{Susceptibility}

The susceptibility can now be obtained from the external field $H_{\sigma}$, Eq. (\ref{Hsigma}), in accordance with the thermodynamic definition
\begin{equation}
\label{su}
\chi^{-1} =\left(\frac{\partial H_{\sigma}}{\partial r} \right)_{T}\,,
\end{equation}
where this now allows us firsthand a better understanding for the response of the oscillator system in relation to the external field behavior, i.e., the response of the system related to the action of multiplicative noise.

Figure 6 shows the isotherms of order parameter $r$ as a function of external field $H_{\sigma}$ Eq. (\ref{Hsigma}). It is remarkable that all curves saturate at $r=1$ to large values of $H_{\sigma}$, analogous to the behavior of magnetization. This concurs with the fact that $H_{\sigma}$ plays the role of an external field, as expected. We can observe that conjugate variables $(r,H_{\sigma})$ have a maximum value for $\sigma=-1$ and decrease toward $(0,0)$ for $\sigma=1$, {\it i.e.}, $T_{eff}=1$, which is the final point of the curve. The isotherm $T=1$ delimits the synchronized phase $T\leq 1$ from the parasynchronized phase $T>1$. Note that the parasynchronized phase exists only for $H_{\sigma}>0$ and its susceptibility is always finite and positive, except for $T=1$ where it diverges as $H_{\sigma} \rightarrow 0$ due to the critical behavior of the system.

In the synchronized phase, for $T=0.7$, as $\sigma$ increases, $H_{\sigma}$ decreases and $r$ decreases, reaching a non-null value at the first zero of $H_{\sigma}(r)=0$, which corresponds to spontaneous synchronization, {\it i.e.}, synchronization in a null field. This behavior is very similar to what happens in magnetization. As $\sigma$ continues to increase, $r$ decreases and $T_{eff}$ increases toward $T_{eff}=1$ for which $r=H_{\sigma}=0$.

The synchronized phase for $H_{\sigma}\leq 0$ exhibits an anomalous behavior since it allows a region with $\chi < 0$ and a divergence of $\chi$. This is because field $H_{\sigma}$ is not a trivial function of $r$ and $\sigma$. It exhibits two zeros, one trivial for $r=0$ and a second for $r=\hat{r}$, where between the two zeros, we have a minimum $(\frac{\partial H_{\sigma}}{\partial r})|_{r=r^{*}}=0$. Consequently, for $0<r<r^{*}$, susceptibility is negative. Indeed, a similar situation occurs in complex liquids such as water where density decreases with temperature in the region $0 ^{\circ}C \leq T \leq 4 ^{\circ}C$.

\begin{figure}[!h]
\label{fig6}
\centering\rotatebox{-90}{\resizebox{7cm}{!}{\includegraphics{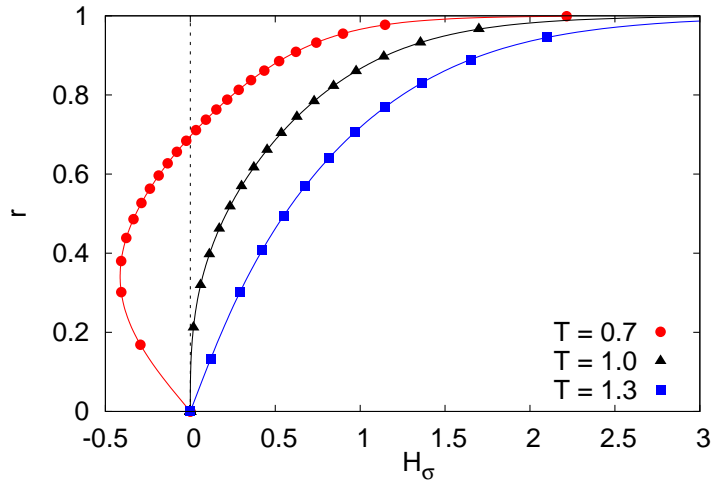}}}
\caption{\label{fig6} Order parameter $r$ as a function of external field $H_{\sigma}$ for several isotherms $T$. All curves saturate at $r=1$, analogous to a magnetic system. The curve $T=1$ separates the synchronized phase $T\leq 1$ from the parasynchronized phase $T>1$.}
\end{figure}

\begin{figure}[!h]
\label{fig7}
\centering\rotatebox{-90}{\resizebox{7cm}{!}{\includegraphics{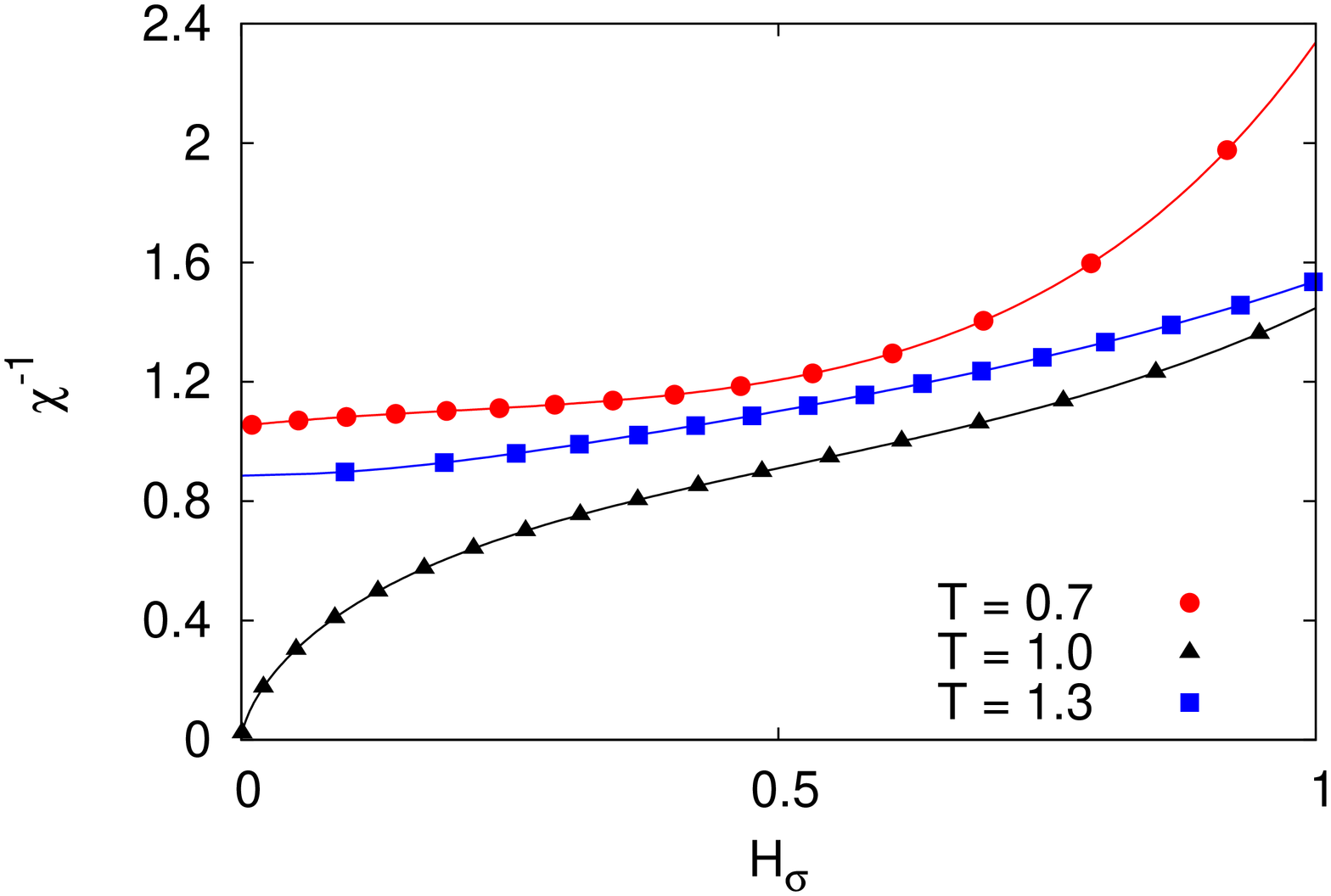}}}
\caption{\label{fig7} Inverse of susceptibility $\chi^{-1}$ as a function of external field $H_{\sigma}$ for fixed isotherms $T$. Curves $T=1.3$ and $T=1$ correspond to the parasynchronized phase. Curve $T=0.7$ corresponds to the synchronized phase. Note that for critical temperature $T=1$ and null field $H_{\sigma}\rightarrow 0$, $\chi=(T_{c}-T)^{-1}$ diverges, as expected.}
\end{figure}

For a liquid, consider the mathematical identity
 \begin{equation}
\label{anomaliaH2O}
\left(\frac{\partial p }{ \partial V} \right)_{T} \left( \frac{\partial T }{ \partial p} \right)_{V} \left( \frac{\partial V }{ \partial T} \right)_{p}=-1\,,
\end{equation}
where $p$ is the pressure and $V$ is the volume. Since the isothermal compressibility
\begin{equation}
\kappa_{T}=-\frac{1}{V}\left(\frac{\partial V }{ \partial p} \right)_{T}\,,
\end{equation}
is always positive, $\kappa_{T}>0$, then in the critical region
\begin{equation}
\left( \frac{\partial V }{ \partial T} \right)_{p}<0\quad\mbox{implies}\quad\left( \frac{\partial T }{ \partial p} \right)_{V}<0\,,
\end{equation}
which satisfies Eq. (\ref{anomaliaH2O}). Mathematically, this is easily understood because $p(V,T)$ is not a linear function of $V$; consequently, it has extremes with null derivatives. The physical mechanism is more complicated but well studied in the literature \cite{Barbosa1,Barbosa2}.

For our system, consider a similar identity
\begin{equation}
\label{anomalia}
\left( \frac{\partial H_{\sigma} }{ \partial r} \right)_{T} \left( \frac{\partial T }{ \partial H_{\sigma}} \right)_{r} \left( \frac{\partial r }{ \partial T} \right)_{H_{\sigma}}=-1\,.
\end{equation}
Since $\left( \frac{\partial T }{ \partial H_{\sigma}} \right)_{r} $ is always positive,
negative susceptibility $\chi=\left( \frac{\partial r}{ \partial H_{\sigma}} \right)_{T}<0$ in the anomalous region implies $\left( \frac{\partial r }{ \partial T} \right)_{H_{\sigma}} >0$. This is very similar to the liquid anomaly.

Figure 7 shows the inverse of susceptibility $ \chi^{-1}$ as function of field $H_{\sigma}$. The temperatures are $T=1.3$ and $T=1$ for the parasynchronized phase and $T=0.7$ for the synchronized phase. As $ H_{\sigma} \rightarrow 0$, the susceptibility $\chi$ tends to a finite value, except for the curve $T=1$ where it diverges, as expected.

\section{Conclusions}

In this article, we presented the full thermodynamics of phase synchronization for a system governed by an internal multiplicative noise. Our starting point is the first law of thermodynamics where we determine the free energy, entropy, internal energy, specific heat, and a synchronization field for the phase oscillator system. From entropy and specific heat, we show that for low temperatures, the synchronized state behaves in very similar fashion to the state of an ideal gas. The synchronization field can be decomposed as $H_{s}=2r+H_{\sigma}$, which plays the role of an effective field for which we define an internal field $H_{0}=2r$ and external field $H_{\sigma}$. The internal field $H_{0}$ corresponds to the bare Gaussian white noise effect on the system, while the external field $H_{\sigma}$ is associated with the phase-dependence effect on the oscillator system, i.e., related to the multiplicative noise action on synchronization. We should remark that susceptibility, defined from the synchronization field, relates to the response of the system due to the multiplicative noise action. From susceptibility, we show that the synchronized phase exhibits anomalous behavior that is analogous to complex liquids such as water. This is a topic that deserves further study in a forthcoming article.

Altogether this system displays a rich behavior featuring characteristics of the magnetic system, classical ideal gas, and anomalies of complex liquids. Interesting systems such as genetic networks \cite{Luonan} and neural systems \cite{Plentz} may also constitute rich fields for the application of concepts such as those we have developed. In particular, we can highlight a potential application in emerging studies of neuronal avalanches on the basis of the Ising model \cite{Yu}, where thermodynamic quantities such as susceptibility and specific heat have been used to describe the physical behavior of these systems. In fact, this phenomenon can be also extensively studied using the thermodynamics of phase oscillators, as presented in this article.

\section{Acknowledgments}

 We acknowledge the support of CNPq, CAPES, and FAP-DF. FAO would like to thank professor Hyunggyu Park for his hospitality during my visit at KIAS.

\appendix
\section*{Appendix A: Null current from stationary density}
\setcounter{section}{1}

We can now demonstrate that the stationary phase distribution $\rho_{s}$ (\ref{hypermises}) results in a null current $J=0$, satisfying the thermodynamic equilibrium criterion \cite{Lax}. We can apply the ergodic analysis for mean frequency $\Omega$ of the oscillator system as
\begin{equation}
\Omega=\lim_{t\rightarrow\infty}t^{-1}\int^{t}_{0}\dot{\theta_i}dt\,.
\end{equation}
This expression can be calculated by replacing the time average with the ensemble average frequency, {\it i.e.} $\Omega=<\dot{\theta_i}>$. Using the Langevin Eq. (\ref{eq:lang_mult}), this results in
\begin{equation}
\Omega=<\dot{\theta_i}> = <rK\sin(\psi -\theta_i)>=\int^{2\pi}_{0}rK\sin(\psi -\theta_i)\rho_{s}(\theta)d\theta \nonumber,
\end{equation}
where $<\sqrt{1+r\sigma\cos(\psi - \theta_i)}\xi_i(t)>=<\sqrt{1+r\sigma\cos(\psi - \theta_i)}><\xi_i(t)>=0$ (there is no correlation between $\theta$ and $\xi$). We can now write the Fokker--Planck equation in terms of current density $J$ as
\begin{equation}
\label{fokker2}
\frac{\partial\rho}{\partial t}=-\frac{\partial J(\theta,t)}{\partial\theta}.
\end{equation}
Hence, comparing (\ref{fokker2}) with (\ref{fokkerg}), we get
\begin{equation}
J(\theta,t)=rK\sin(\psi -\theta_i)\rho(\theta,t)-D\frac{\partial}{\partial\theta}\big[(1+r\sigma\cos(\psi - \theta_i))\rho(\theta,t)\big] \nonumber.
\end{equation}
Note that the condition for the stationary solution is $t\rightarrow\infty$ on (\ref{fokker2}), where $\rho(\theta,\infty)=\rho_{s}(\theta)$ corresponds to $J(\theta,\infty)=J_{s}=$constant. Now using periodic boundary condition $\rho_{s}(\theta+2\pi)=\rho_{s}(\theta)$, the expression for $\Omega$ is given by
\begin{eqnarray}
\Omega &=& \int^{2\pi}_{0}J_{s}d\theta +D\int^{2\pi}_{0}\frac{\partial}{\partial\theta}\big[(1+r\sigma\cos(\psi - \theta_i))\rho_{s}\big]d\theta. \nonumber\\
\nonumber\\
&=& 2\pi J_{s}+D\big[(1+r\sigma\cos(\psi -2\pi))\rho_{s}(2\pi)-(1+r\sigma\cos(\psi))\rho_{s}(0)\big]= 2\pi J_{s}\,\,. \nonumber
\end{eqnarray}
Therefore, we can express $J_{s}$ as
\begin{equation}
J_{s}=(K-D\sigma)r\sin(\psi -\theta_i)\rho_{s}-D\big[(1+r\sigma\cos(\psi - \theta_i))\frac{\partial\rho_{s}}{\partial\theta}\big] \nonumber.
\end{equation}
Taking the derivative $\partial\rho_{s}/\partial\theta$ from Eq. (\ref{hypermises})
\begin{equation}
\frac{\partial\rho_{s}}{\partial\theta}=\frac{\nu\sigma r\sin(\psi-\theta)\rho_{s}(\theta)}{1+r\sigma\cos(\psi - \theta_i)}\,,
\end{equation}
Finally, we can write
\begin{equation}
\label{JJ}
J_{s}=\big[K-D\sigma(1+\nu)\big]r\sin(\psi -\theta_i)\rho_{s}\,.
\end{equation}
However, we know that $\nu=K/D\sigma-1$, which immediately implies $J_{s}=0$. Then the stationary phase density $\rho_s$ directly results in a null current $J_s$ in the system, which is the condition for the thermodynamic equilibrium.

\section*{Appendix B: Order parameter}
\setcounter{section}{2}

To calculate the order parameter, we use the Legendre functions defined in Gradshteyn \cite{Gradshteyn} (8.711.2), given by
\begin{eqnarray}
\label{order2g}
\!\!\!\!\!\!\!\!\!\!\!\!\!\!P^{m}_{\nu}(z)&=&\frac{(\nu+1)(\nu+2)\cdots(\nu+m)}{\pi}\int^{\pi}_{0}\big[z+\sqrt{z^2-1}\cos\varphi\big]^{\nu}\cos m\varphi d\varphi\\
\nonumber\\
&=&\frac{(-1)^{m}\nu(\nu-1)\cdots(\nu-m+1)}{\pi}\int^{\pi}_{0}\frac{\cos m\varphi d\varphi}{\big[z+\sqrt{z^2-1}\cos\varphi\big]^{\nu+1}}\,.
\end{eqnarray}

To calculate the critical coupling $K_{c}$, we take the Legendre functions in asymptotic forms ~\cite{Erdelyi}
\begin{equation}
P^{1}_{\nu}(z)\sim\frac{\Gamma(\nu+2)}{\sqrt{2}\Gamma(\nu)}(z-1)^{1/2}\quad\mbox{and}\quad P_{\nu}^{0}(z)\sim 1,
\end{equation}
for $z\approx 1$ (see Eq. (\ref{constn})), where $\Gamma(x)$ is the Gamma function. Then taking the expansion of $z$ as $r\approx 0$ and inserting it in the asymptotic limit of Eq. (\ref{order2}), we obtain
\begin{equation}
\label{BK}
r=\frac{\nu\sgn(\sigma)}{\sqrt{2}}\left(\frac{r^{2}\sigma^2}{2}\right)^{1/2}=\frac{\nu r\sigma}{2}.
\end{equation}


\section*{References}
\begin{thebibliography}{99}

\bibitem{Strogatz} S. Strogatz, {\it SYNC: The Emerging Science of Spontaneous Order} (Hyperion, New York, 2003).

\bibitem{Longa} L. Longa, E. M. F. Curado, F. A. Oliveira, Phys. Rev. E {\bf 54}, 2201 (1996).

\bibitem{Ciesta} M. Ciesla, S. P. Dias, L. Longa, and F. A. Oliveira, Phys. Rev. E {\bf 63}, 065202 (2001).

\bibitem{Morgado} R. Morgado, M. Ciesla, L. Longa, F. A. Oliveira, Europhys. Lett. {\bf 79}, 10002 (2007).

\bibitem{Acebron} J. A. Acebron, L. L. Bonilla, C. J. Perez Vicente, F. Ritort, R. Spigler, Rev. Mod. Phys. {\bf 77}, 137 (2005).

\bibitem{Hong} H. Hong, H. Park, L. Tang, Phys. Rev. E {\bf 76}, 066104 (2007).

\bibitem{Bonilla} L. L. Bonilla, C. J. P\'erez Vicente, J. M. Rub\'i, J. Stat. Phys. {\bf 70}, 921 (1993).

\bibitem{Sonnenschein} B. Sonnenschein, L. Schimansky-Geier, Phys. Rev. E {\bf 88}, 052111 (2013).

\bibitem{Park} S. H. Park, S. Kim, Phys. Rev. E {\bf 53}, 3425 (1996).

\bibitem{Reimann} P. Reimann, C. Van den Broeck, R. Kawai, Phys. Rev. E {\bf 60}, 6402 (1999).

\bibitem{Yu} S. Yu, H. Yang, O. Shriki, D. Plenz, Front. Syst. Neurosci. {\bf 7}, 42 (2013).

\bibitem{Kuramoto1} Y. Kuramoto, Chemical Oscillations, Waves, and Turbulence (Springer-Verlag, Berlin, 1984).

\bibitem{Nakao} H. Nakao, K. Arai, Y. Kawamura, Phys. Rev. Lett. {\bf 98}, 184101 (2007).

\bibitem{Zhou} C. Zhou, J. Kurths, Phys. Rev. Lett. {\bf 88}, 230602 (2002).

\bibitem{Teramae} J. Teramae, D. Tanaka, Phys. Rev. Lett. {\bf 93}, 204103 (2004).

\bibitem{Pikovsky2} D.S. Goldobin, A. Pikovsky, Phys. Rev. E. {\bf 71}, 045201(R) (2005).

\bibitem{Pikovsky} A. Pikovsky, M. Rosenblum, J. Kurths, {\it Synchronization: A Universal Concept in Nonlinear Sciences} (Cambridge University Press, Cambridge, 2001).

\bibitem{Nagai} K. H. Nagai, H. Kori, Phys. Rev. E {\bf 81}, 065202 (2010).

\bibitem{Lai} Y. M. Lai, M. A. Porter, Phys. Rev. E {\bf 88}, 012905 (2013).

\bibitem{Hanggi} G. Schmid, P. Hanggi, Mathematical Biosciences {\bf 207}, 235 (2007).

\bibitem{Crawford} J. D. Crawford, Phys. Rev. Lett. {\bf 74}, 4341 (1995).

\bibitem{Gupta} S. Gupta, A. Campa, S. Ruffo, J. Stat. Mech., R08001 (2014).

\bibitem{Sasa} Shin-ichi Sasa, New J. Phys. {\bf 17}, 045024 (2015).

\bibitem{Sakaguchi} H. Sakagushi, Prog. Theor. Phys. {\bf 79}, 39 (1988).

\bibitem{Bonilla2} L.L. Bonilla, C.J. P\'erez Vicente, F. Ritort, J. Soler, Phys. Rev. Lett. {\bf 81}, 3643 (1998).

\bibitem{Gardiner} C. W. Gardiner, {\it Handbook of Stochastic Methods} Four edition (Springer-Verlag, 2009).

\bibitem{Yoshimura} K. Yoshimura, K. Arai, Phys. Rev. Lett. {\bf 101}, 154101 (2008).

\bibitem{Teramae2} J. Teramae, H. Nakao, G.B. Ermentrout, Phys. Rev. Lett. {\bf 102}, 194102 (2009).

\bibitem{Risken} H. Risken, {\it The Fokker-Planck equation} 2nd edition (Berlin: Springer-Verlag, 1996).

\bibitem{Russel} D. I. Russell, R. A. Blythe, J. Stat. Mech., P06008 (2013).

\bibitem{Hongler} M. O. Hongler, R. Filliger, P. Blanchard, Europhys. Lett. {\bf 89}, 10001 (2010).

\bibitem{Ran} G. Ran, D. Jiulin, Physica A {\bf 406}, 281 (2014).

\bibitem{Kampen} N. G. van Kampen, J. Stat. Phys. {\bf 24}, 175 (1981).

\bibitem{Kuroiwa} T. Kuroiwa, K. Miyazaki, J. Phys. A: Math. Theor. {\bf 47}, 012001 (2014).

\bibitem{Lau} A. W. C. Lau, T. C. Lubensky, Phys. Rev. E {\bf 76}, 011123 (2007).

\bibitem{Barbosa1} M. A. Barbosa, F. V. Barbosa, F. A. Oliveira, J. Chem. Phys. {\bf 134}, 024511 (2011).

\bibitem{Barbosa2} M. A. Barbosa, E. Salcedo, M. C. Barbosa, Phys. Rev. E {\bf 87}, 032303 (2013).

\bibitem{Basnarkov} L. Basnarkov, V. Urumov, Phys. Rev. E {\bf 76}, 057201 (2007).

\bibitem{Landau} L. D. Landau, E. M. Lifshitz, {\it Statistical Physics} Vol. 5 (Oxford: Pergamon Press, 1980).

\bibitem{Falk} H. Falk, L. W. Bruch, Phys. Rev. {\bf 180}, 442 (1969).

\bibitem{Luonan} L. Chen et al, {\it Modelling Biomolecular Networks in Cells} (Springer: London, 2010).

\bibitem{Plentz} D. Plenz, E. Niebur, H. G. Schuster, {\it Criticality in Neural Systems} (Wiley-VCH Verlag: Weinheim, Germany, 2014).

\bibitem{Lax} M. Lax, Rev. Mod. Phys. {\bf 38}, 359 (1966).

\bibitem{Gradshteyn} I. S. Gradshteyn, I. M. Ryzhik, {\it Table of Integrals, Series, and Products} 7th edition (Academic Press, New York, 2007).

\bibitem{Cohl} H. S. Cohl, J. Classical Anal. {\bf 2}, 107 (2013).

\bibitem{Erdelyi} A. Erdelyi et al., {\it Higher Transcendental Functions} Vol. I (McGraw-Hill, New York, 1953).

\bibitem{Szmytkowski} R. Szmytkowski, J. Phys. A: Math. Gen. {\bf 39}, 15147 (2006).

\bibitem{Olver} F. W. J. Olver, {\it Asymptotics and Special Functions} (Academic Press, New York, 1974).







\end{thebibliography}
\end{document}